\pdfoutput=1
\documentclass[12pt]{article}
\usepackage[margin=0.8in]{geometry}
\usepackage{natbib}

\usepackage{comment}
\usepackage[percent]{overpic}
\usepackage[flushleft]{threeparttable}
\usepackage[clockwise, figuresright]{rotating}
\usepackage{pbox}
\usepackage{adjustbox}
\usepackage{multirow}
\usepackage{longtable}
\usepackage{enumitem}
\usepackage{rotating} 
\usepackage{array}
\newcolumntype{P}[1]{>{\centering\arraybackslash}p{#1}}
\usepackage{setspace}
\usepackage{authblk}
\usepackage[colorlinks=true,linkcolor=blue,allcolors=blue]{hyperref}   
\usepackage{url}  

\def\soho{{\sl SOHO}}
\def\stereo{{\sl STEREO}}

\def\apj{Astrophys~J~}
\def\apjl{Astrophys~J~Lett~}

\def\ssr{Spac~Sci~Rev~}

\def\apjs{\apj Suppl~}

\def\solphys{Sol~Phys~}

\def\aap{Astro~and~Astrophys~}

\begin{document}

\title{Magnetohydrodynamic Simulation of a {Coronal Mass Ejection} Observed During the Near-radial Alignment of Solar Orbiter and Earth}

\author[1]{Talwinder Singh}
\affil[1]{Department of Physics \& Astronomy, Georgia State University, Atlanta, GA 30303, USA}

\author[2]{Dinesha V. Hegde}
\affil[2]{Department of Space Science, The University of Alabama in Huntsville, AL 35805, USA}

\author[3]{Tae K. Kim}
\affil[3]{Center for Space Plasma and Aeronomic Research, The University of Alabama in Huntsville, AL 35805, USA}

\author[2,3]{Nikolai V. Pogorelov}

\maketitle

\begin{abstract}
{Interplanetary }Coronal Mass Ejections {(ICMEs)} are the primary sources of geomagnetic storms at Earth. Negative out-of-ecliptic component (Bz) of magnetic field {in the ICME or its associated sheath region} is necessary for it to be geo-effective. For this reason, magnetohydrodynamic simulations of CMEs containing data-constrained flux ropes are more suitable for forecasting their geo-effectiveness {as compared to hydrodynamic models of the CME}. ICMEs observed {in situ} by radially aligned spacecraft can provide an important setup to {validate the physics-based heliospheric modeling of CMEs}. In this work, we use the constant-turn flux rope (CTFR) model to study a{n I}CME that was observed in situ by Solar Orbiter (SolO) and at Earth, when they were in a near-radial alignment. This was a stealth CME that erupted on {2020 April 14} and reached Earth on {2020 April 20} with a weak shock and a {smoothly rotating magnetic field} signature. We found that the CTFR model was able to reproduce the {rotating magnetic field} signature at both SolO and Earth with very good accuracy. The simulated {I}CME arrived 5 hours late at SolO and 5 hours ahead at Earth, when compared to the observed {I}CME. We compare the propagation of the CME front through the inner heliosphere using synthetic J-maps and those observed in the heliospheric imager data and discuss the role of incorrect ambient SW background on kinematics of the simulated CME. This study supports the choice of the CTFR model for {reproducing the} magnetic field of {I}CMEs.
\end{abstract}

Coronal Mass Ejections --- Graduated Cylindrical Shell Model --- Magneto-hydrodynamics --- Space Weather Prediction

\section{Introduction}\label{sec:Introduction}
Coronal mass ejections (CMEs) are structures {erupting from the solar corona and} composed of plasma and magnetic field, typically originating at high {magnetic field} strength, closed-field regions of the Sun known as active regions (ARs). CMEs are capable of transporting substantial kinetic and magnetic energy {from the low corona out to the heliosphere}, with the erupted mass varying from \(10^{9}\) kg to \(10^{13}\) kg \citep{Vourlidas10} and speeds occasionally surpassing 3000 km s$^{-1}$. {CMEs, when observed in situ are referred to as interplanetary CMEs \citep[ICMEs;][]{Rouillard11}}. An Earth-bound {I}CME with a negative ({south-ward}) out-of-ecliptic magnetic field component (\(B_\mathrm{z}\)) has a potential to trigger severe space weather {conditions} impacting both space-based and terrestrial technological systems {\citep{Howard11,Webb12}}. Consequently, accurate  forecasts of the arrival time and magnetic field characteristics of {I}CMEs at Earth, and description of their interplanetary transit are  of critical importance {for mitigating their harmful effects}. 

{The CME plasma travels together with an expanding magnetic field, which usually presents itself as having helical field lines. This structure is called a magnetic flux rope \citep{Webb12}.} The treatment of CME flux ropes plays a critical role in developing forecast models that predict the magnetic properties of {I}CMEs at 1 au \citep{Gopalswamy2018a,Sarkar2020}. Consequently, a new generation of forecast models {employ a flux rope model to characterize the magnetic structure of the ejectas} \citep{Shen11, Shiota16, Isavnin16, Vandas17, Jin17b, Scolini19, Singh20b, Singh22}. {These approaches model CMEs by inserting a flux rope magnetic field and appropriate plasma conditions into the} ambient solar wind (SW) {constructed based on} boundary conditions (BCs) at the Sun. 

{The flux rope modeling as a forecasting tool can be broadly divided into two categories: empirical and physics-based. Empirical modeling involves defining the flux rope as an idealized shape and expanding it in the corona and inner heliosphere by assuming a simplified interaction with the ambient SW \citep[e.g.][]{Kay16,Weiss21}. The physics based modeling employs magnetohydrodynamic (MHD) simulations to capture the expansion of flux ropes and their interaction with the SW. In these simulations, the flux rope evolution either starts close to the sun \citep[e.g.][]{Jin16,Singh18,Lynch21} or above the critical surface at around 0.1 au \citep[e.g.][]{Shen14, Singh20b, Maharana22, Palmerio23, Mayank24}. The latter approach involves assumptions like self-similar expansion of the flux rope in the corona, but it is more suitable for operational purposes since coronal MHD modeling can be computationally expensive.}

\citet{Singh22} proposed a constant-turn flux rope (CTFR) model that uses {a global} geometry similar to the FRiED model of \citet{Isavnin16} and specifies {the} magnetic field in it using the uniform turn solution proposed by \citet{Vandas17}. Using this model in an  {inner heliosphere} MHD simulation, the magnetic field profile of the {2012 July 12} CME was shown to be reproduced at Earth with good accuracy. In this work, we use the CTFR model to simulate a stealth CME. The CME erupted from the Sun on {2020 April 14} but did not exhibit any {readily apparent} eruption signatures in the extreme ultraviolet (EUV) observations in the lower corona, thus falling into the category of stealth CMEs. {The CME was observable by the STEREO and SOHO coronagraphs, making this CME not a ``super''-stealth CME \citep{Nitta21}, and thus the geomagnetic storm caused by this CME cannot be categorized as a ``problem'' geomagnetic storm, which are the storms caused by CMEs not observed in both EUV and coronagraph data.} This CME, travelling through the inner heliosphere passed over Solar Orbiter (SolO) and Earth when they were separated by 0.19 au in radial direction and less than 2$^\circ$ in longitude. This was a slow {I}CME with a weak shock observed in front of it at SolO and at Earth. The magnetic field configuration showed a smoothly rotating profile in both measurements. Therefore, the radially-aligned location of SolO and Earth together with the uncomplicated magnetic field signature of the {I}CME provided a rare opportunity to verify the CTFR model by simulating the radial evolution of CMEs. The results presented in this study suggest that the CTFR model is a good candidate to improve forecasting of the magnetic field properties of {I}CMEs, especially when a{n I}CME is characterized by a simple {magnetic field} structure.

In Section~\ref{sec:Data}, we describe the data used in this work. In Section~\ref{sec:Models}, we outline our ambient SW and CME models. This is followed by the description of our simulations and their analysis in Section~\ref{sec:Results}. The the conclusions are presented in Section~\ref{sec:Conclusions}.

\section{Data Used}\label{sec:Data}
In this work, we examine a stealth CME that occurred on {2020 April 14} and reached Earth on {2020 April 20}. To analyze this CME, we utilize level~0.5 FITS data from the Sun-Earth Connection Coronal and Heliospheric Investigation (SECCHI)/Cor2 coronagraph \citep{Howard08} on the Solar Terrestrial Relations Observatory (\stereo) A \citep{Kaiser08}. This data is processed to level~1 using the \textit{secchi\_prep} function in the IDL SolarSoft library, which converts the data from data numbers (DN) to Mean Solar Brightness (MSB). We also use level~1 FITS data from the Solar and Heliospheric Observatory (\soho) spacecraft's Large Angle Spectroscopic Coronagraph (LASCO)/C2~\citep{Brueckner95} coronagraph. We employ the coronagraph data to determine the CME direction, tilt, half-angle, aspect ratio, and velocity with the graduated cylindrical shell model \citep[GCS][]{Thernisien09}. To track the CME in the inner heliosphere (IH), we use the long-term, background-subtracted data from HI 1 \& 2 \citep{Eyles09} on board \stereo\ A.  Level 2 HI data, already corrected for cosmic rays, shutterless readout, saturation effects, flat fields, and spacecraft pointing offsets, are used in our analysis. To compare in situ measurements at Earth with our simulations, we employ 1-minute averaged plasma and magnetic field data provided by NASA/GSFC's OMNI data via OMNIWeb \citep{King05}, and 1-minute averaged magnetic field data from SolO, available through CDAWeb (\url{https://cdaweb.gsfc.nasa.gov/}).

\section{Simulation Models}\label{sec:Models}
This work employs the Multi-Scale Fluid-Kinetic Simulation Suite (MS-FLUKSS) to perform MHD simulations of the {inner heliosphere}. MS-FLUKSS is a collection of modules capable of performing simulations on adaptive meshes in the presence of neutral atoms, non-thermal ions, turbulence, etc. \citep{Pogorelov14, Fraternale21,Bera2023}. High level of parallelization implemented in this suite enables efficient simulations with the execution faster than real time. We describe the SW and CME models used in this study in the subsequent sections.

\subsection{Solar wind model}
In MS-FLUKSS, the IH model follows the finite-volume, total variation diminishing (TVD) methodology to solve ideal magnetohydrodynamic (MHD) equations on a spherical grid. {The magnetic divergence cleaning is performed using the 8-wave method described by \cite{Powell99}}. The lower boundary conditions for our IH model are derived from the time series of {Wang-Sheeley-Arge} (WSA) maps \citep[e.g.,][]{Kim19}. The BCs for the WSA coronal model are specified at the solar surface and derived from the Air Force Data Assimilative Photospheric Flux Transport (ADAPT) model \citep{Arge10, Arge11, Arge13, Hickmann15}, which assimilated SDO/HMI line-of-sight magnetograms at 12-hour cadence for this study. The WSA model extends the photospheric magnetic field to a spherical source surface at $2.5R_\odot$ using the potential field source surface (PFSS) model. The solution is further extended to the corresponding outer boundary at 0.1 au with the Schatten current sheet model \citep{Schatten71}. The SW speed at this boundary is calculated by WSA based on the magnetic field expansion factor and proximity to the nearest coronal hole boundary \citep{Arge03, Arge05, McGrogor2011} and further uniformly reduced by 75 km s$^{-1}$ to account for the difference in SW acceleration efficiency in the WSA (kinematic) and MHD models \citep[e.g.,][]{Kim19}. The SW density and temperature at 0.1 au are estimated based on the assumptions of constant momentum flux and thermal pressure balance, respectively \citep[e.g.,][]{Linker16}. The IH model employs the WSA output at 0.1 au as boundary conditions \citep{Singh20b,Singh22}. The ADAPT-WSA model generates an ensemble of 12 realizations to be used as the inner BCs for the IH model. Our CME simulations utilize the ADAPT-WSA realization that best reproduces the near-Earth ambient SW for the period of interest.

\subsection{CME model}
For CME simulation in this study, the CTFR model \cite{Singh22} is used. This model uses the FRiED model geometry \citep{Isavnin16} with a croissant-like shape with circular cross-section and twin legs {connected at the origin} as the initial CME shape. The flux rope's magnetic field is derived from {the} analytic solution given by \citet{Vandas17}. The flux rope model allows orientation in any direction, specified by apex latitude and longitude, and permits arbitrary tilt relative to the solar equatorial plane, as well as chosen half-angle and aspect ratio. The magnetic field in the flux rope can have defined poloidal and toroidal fluxes and a helicity sign of either +1 or -1, indicating the {handedness of the twist} following the right- or left-hand rule, respectively. The plasma velocity within the flux rope is initialized with a specific apex value and follows a self-similar expansion profile as detailed in \citet{Singh22}. The option exists to incorporate a uniform-density plasma inside the flux rope. The flux rope is introduced into the IH domain fully formed, i.e., it is not evolving  incrementally by modifications of the inner domain boundary. {The ghost cells below the inner boundary are not altered during CME insertion. At the time of insertion, the legs of the flux rope are in contact with the inner boundary of the domain. As time progresses, the legs move away from this boundary, and the SW magnetic field is introduced in the region beneath them through the inner ghost cells, where the WSA values are prescribed.}

\citet{Singh22} demonstrated this model's suitability for simulating CMEs in the IH, producing realistic magnetic field, density, and velocity results at Earth for the {2012 July 12} CME, using observational data for direction, tilt, half-width, aspect ratio, speed, mass, helicity sign, and magnetic flux. These CME properties can be derived from various observational data, as discussed in \citet{Singh20a}. {However, in the present study involving a stealth CME, observations needed to constrain the magnetic flux of the flux rope are not available. Therefore, we have used assumed values of the magnetic flux.}

\section{Results}\label{sec:Results}

\subsection{Constraining initial CME parameters with data}\label{subsec:Constrain}
\begin{figure}
\centering
\center
\begin{tabular}{c c} 
\includegraphics[scale=0.1,angle=0,width=5.5cm,keepaspectratio]{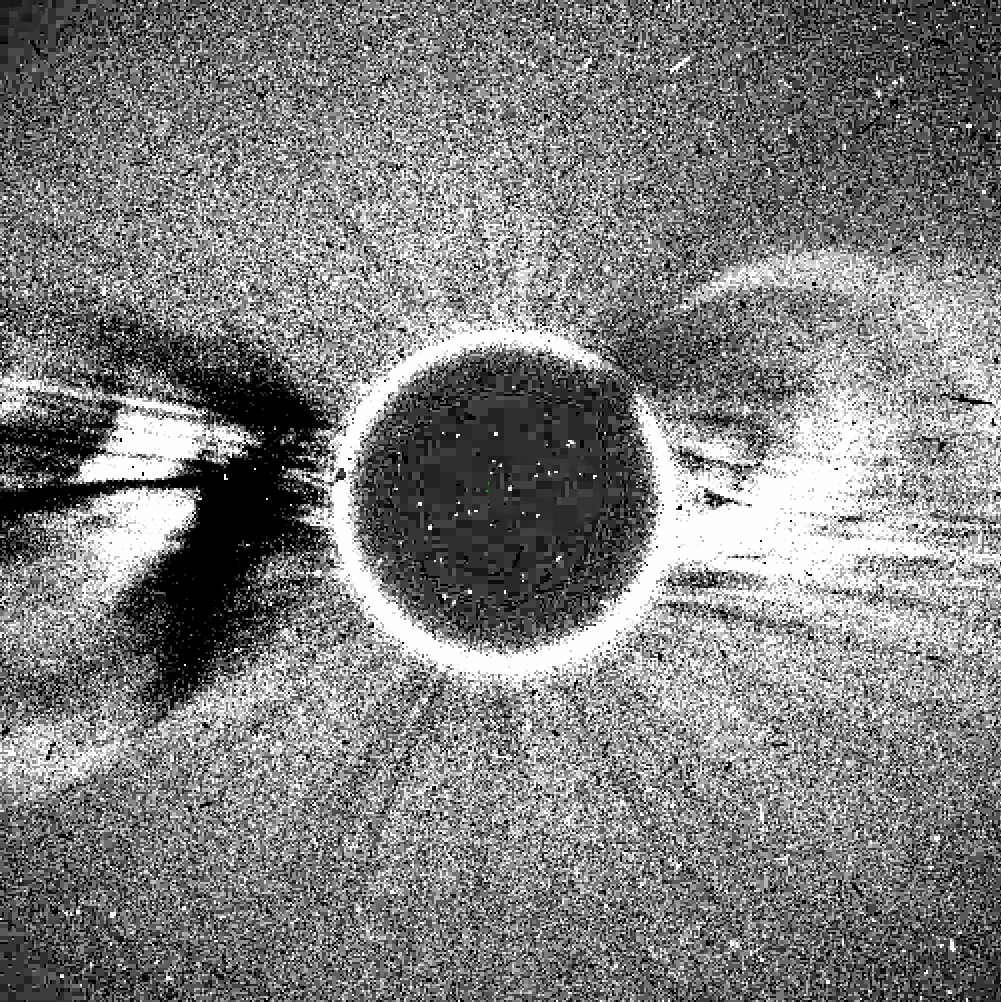}
\includegraphics[scale=0.1,angle=0,width=5.5cm,keepaspectratio]{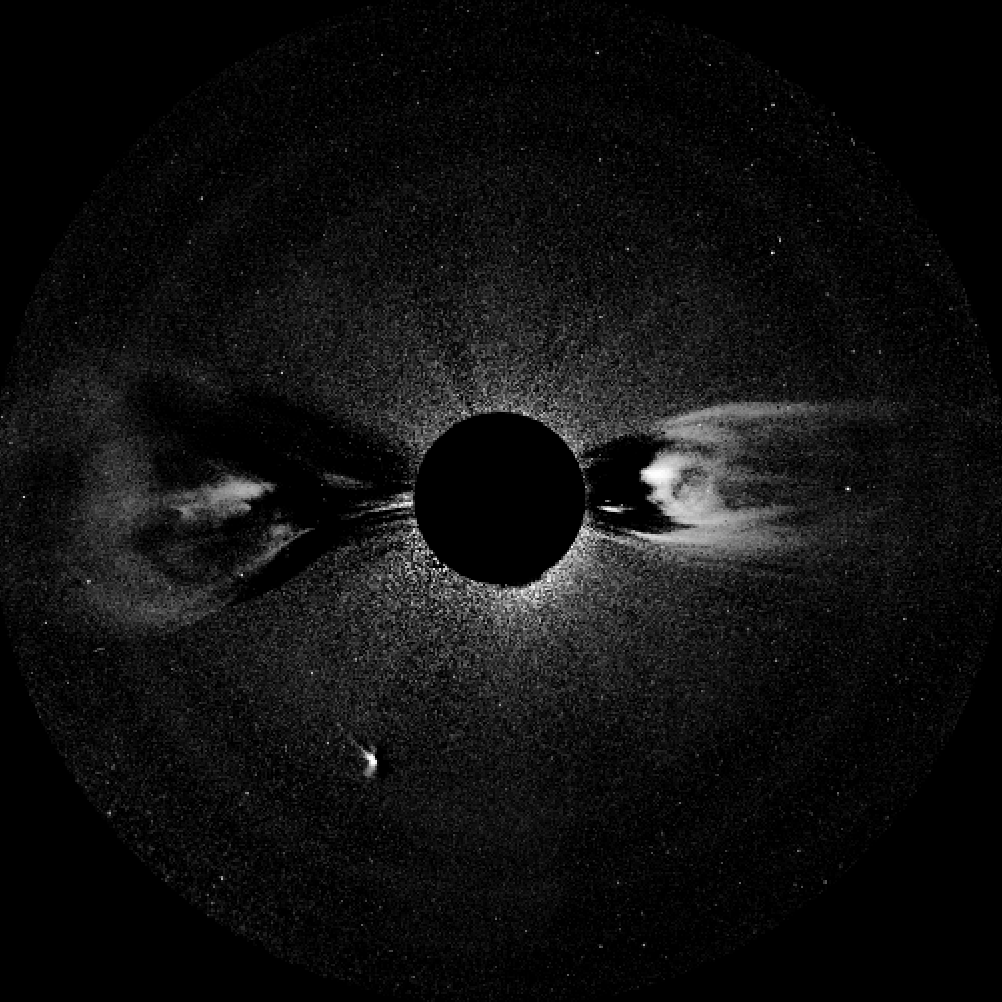}\\
\includegraphics[scale=0.1,angle=0,width=5.5cm,keepaspectratio]{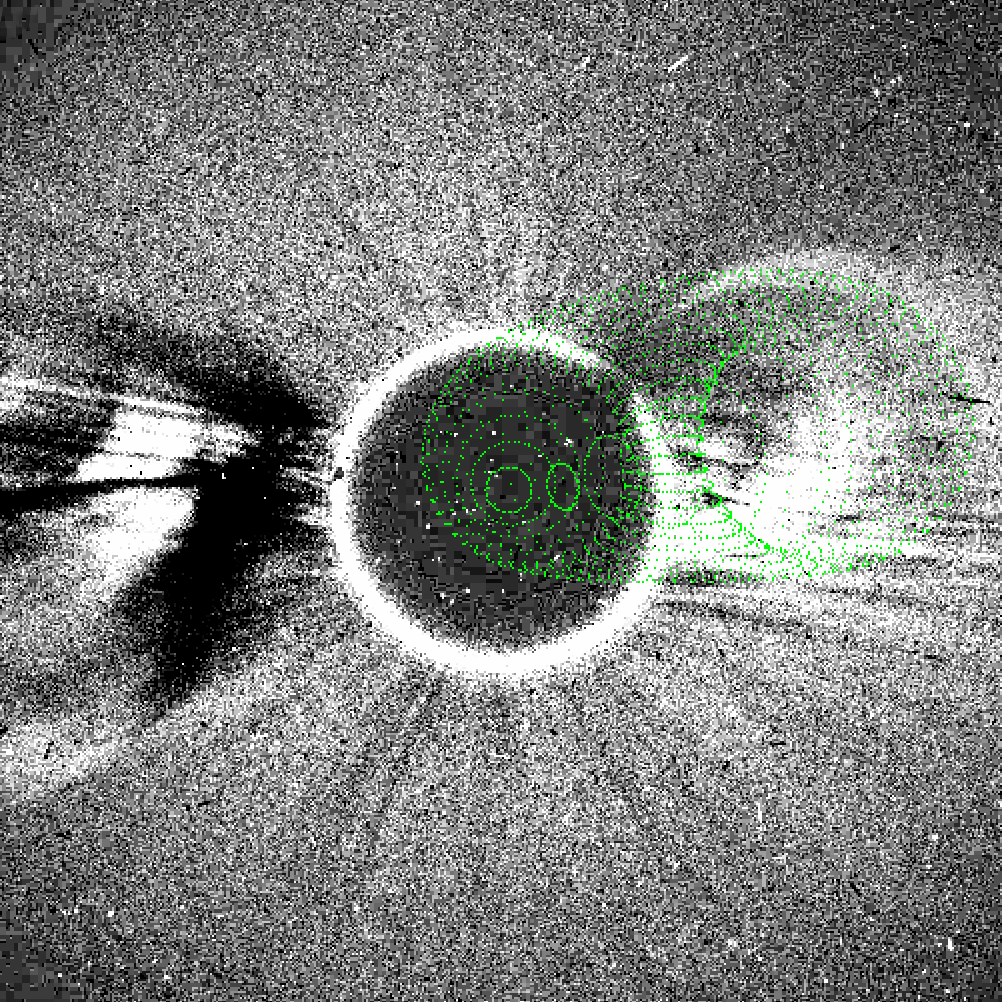}
\includegraphics[scale=0.1,angle=0,width=5.5cm,keepaspectratio]{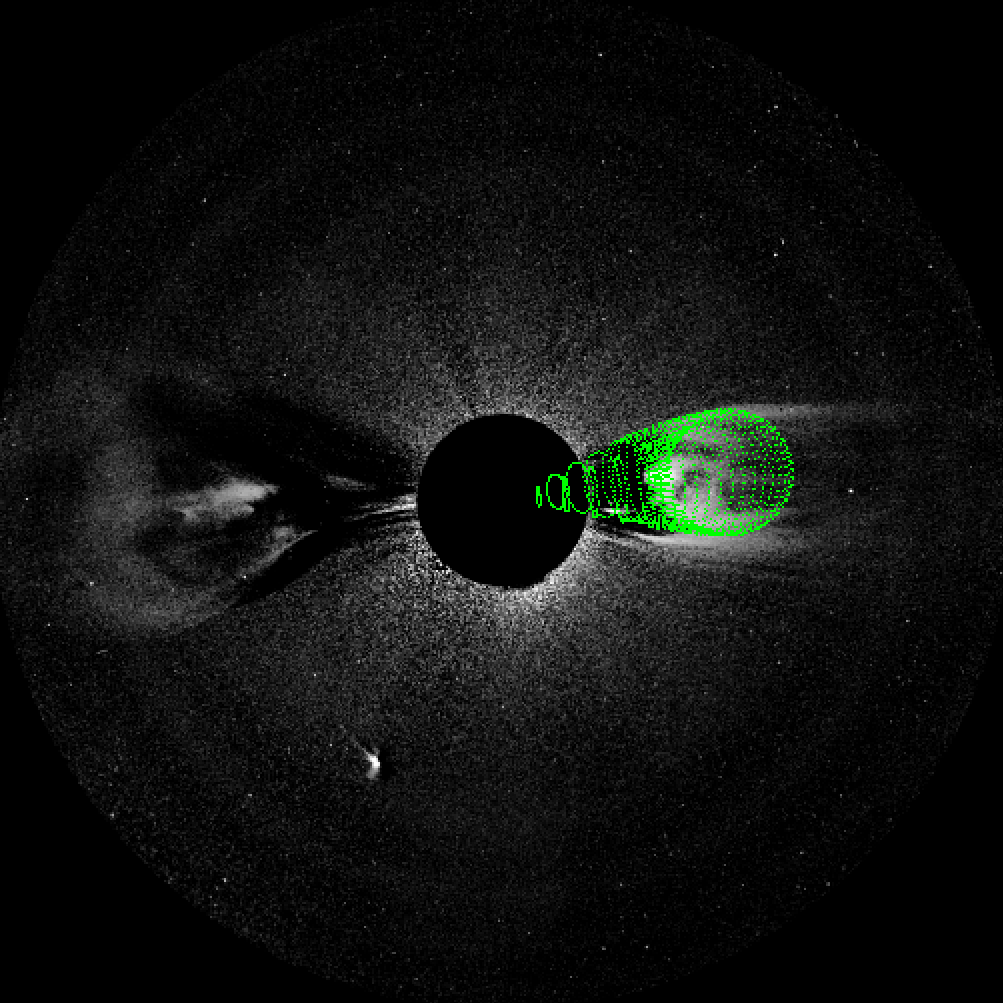}
\end{tabular}
\caption{(\textit{Top panels:}) SOHO LASCO/C2  (\textit{left})  and  STEREO A SECCHI/Cor2 (\textit{right}) coronagraph observations of the stealth CME on {2020 April 15} at 05:24. (\textit{Bottom panels:}) The same as in the top panels, but with GCS fitting overlaid on the CME as a green wired mesh.}
\label{GCS}
\end{figure}

The stealth CME studied in this work did not show any signature near the solar surface. As a result, the magnetic properties of the associated flux rope cannot be determined by the methods that utilize EUV observations such as post eruption arcades (PEAs) and coronal dimming \citep{Gopalswamy17, Dissauer18}. The CME appeared in the STEREO A Cor2 and SOHO C2 coronagraphs on {2020 April 14} as a slowly moving structure. The top panel of figure~\ref{GCS} shows the SOHO C2 and STEREO A Cor2 observations of this CME on {2020 April 15} at 05:24 UT. The bottom panel shows the GCS model fit to this CME. The GCS fitting provides the CME direction, height, tilt, aspect ratio and half angle. Fitting for a time series of images can provide a height-time relation which can be converted to speed via linear regression. We found the CME latitude and longitude to be $4^{\circ}.5$ and $25^{\circ}.5$, {respectively, in {the} Stonyhurst coordinate system. The CME tilt was found to be} $10^{\circ}.1$. The CME speed is 120 km s$^{-1}$. CME half angle and aspect ratio were $21^{\circ}.8$ and $0.27$, respectively. These GCS parameters were used to initialize our CME simulation as discussed in the next section.

\subsection{Simulation Results}\label{sec:Sim_Results}
\begin{figure}
\centering
\includegraphics[scale=0.1,angle=0,width=14cm,keepaspectratio]{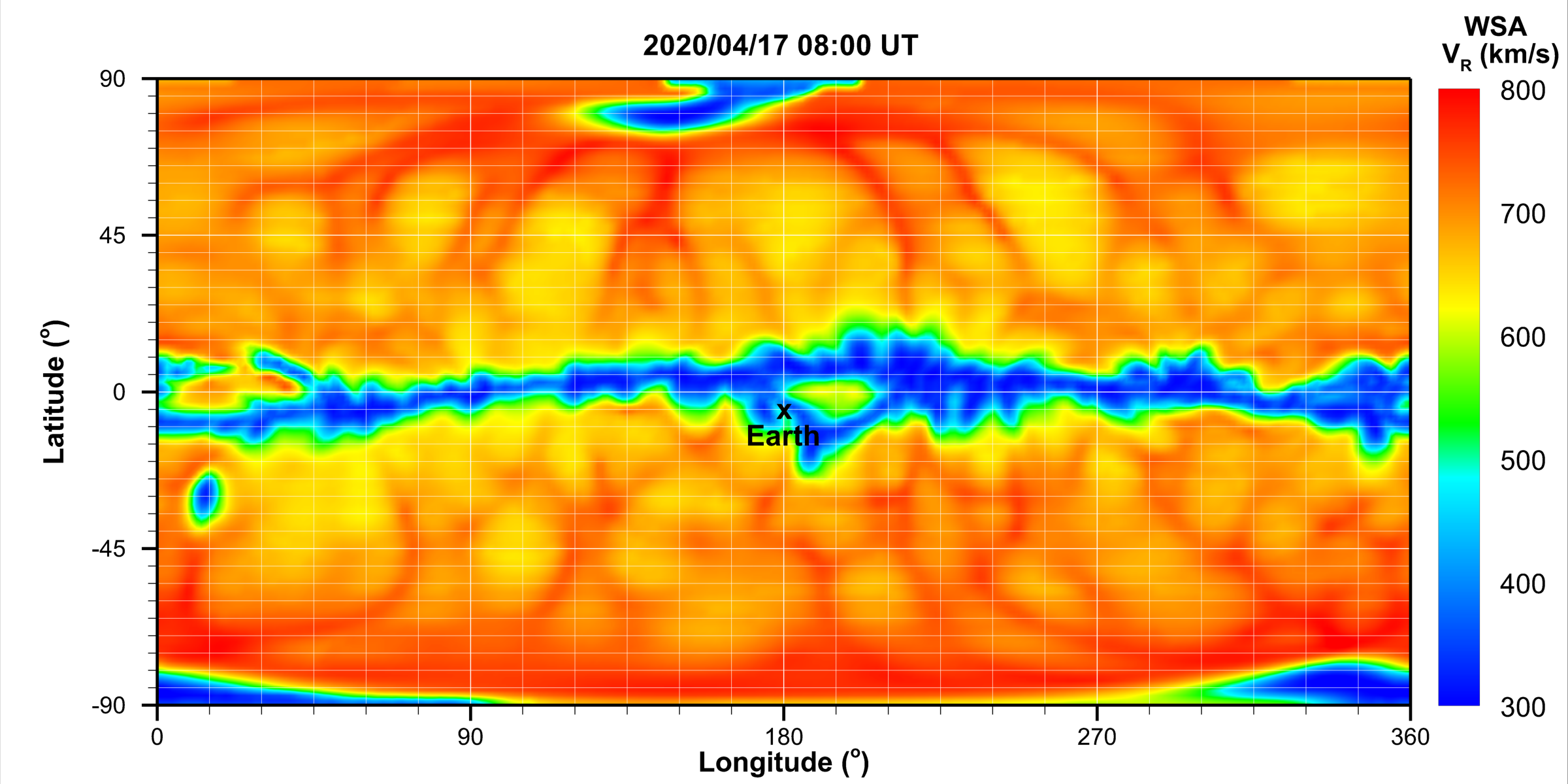}
\includegraphics[scale=0.1,angle=0,width=14cm,keepaspectratio]{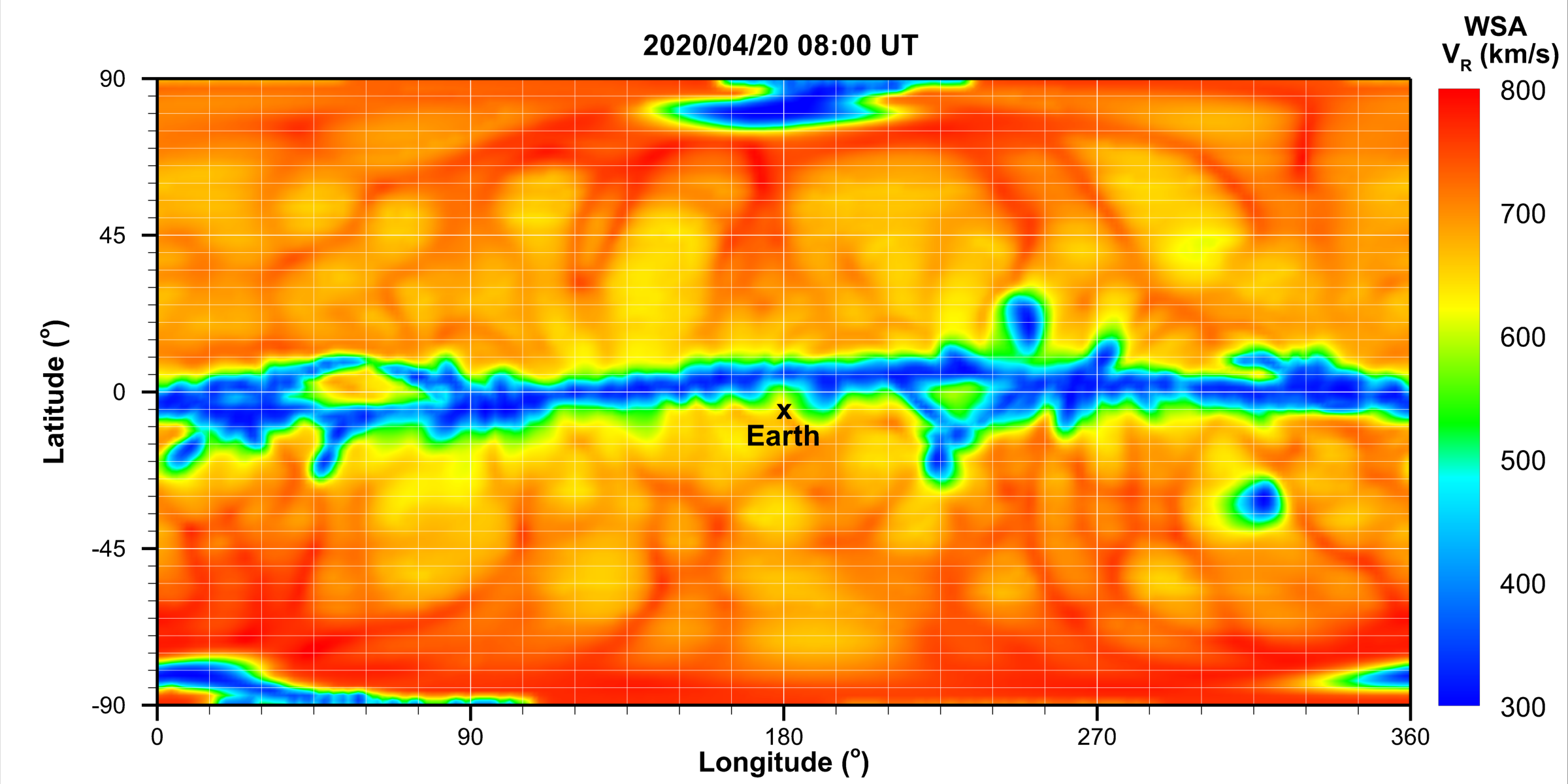}
\caption{Two different frames from the time dependent inner boundary radial velocity maps specified at 0.1 au {at 2020 April 17 08:00 UT (top) and at 2020 April 20 08:00 UT (bottom)}. The vertical and horizontal axes correspond to solar latitudes and Stonyhurst longitudes, respectively. The Earth location is marked with a cross on the central meridian.}
\label{IBC}
\end{figure}

{The s}imulation was performed using MS-FLUKSS on a {fully} spherical grid with non-uniform cell size in {the radial} and {polar} directions. {In {the} radial direction, $r$ varies exponentially with the cell index, resulting in a $dr$ that ensures the conservation of the cell aspect ratio. This is done because the non-radial cell sizes increase linearly with $r$ in spherical geometry. In the {polar} direction, $d\theta$ is kept constant between 8$^\circ$ and 172$^\circ$ polar angles, and increases exponentially near the polar axis. This is done to avoid excessively small cell sizes near the poles and increase the stable time step in accordance with the Courant–-Friedrichs–-Levy (CFL) condition}. In our simulation, a grid of $160 \times 256 \times 128$ ($r \times \phi \times \theta$) resolution was used. The inner boundary of the simulation domain is placed at 0.1 au, where the SW is already {super fast magnetosonic}.  The outer boundary is at 1.5 au. {The inner boundary conditions are derived using the WSA model, which used a time series of ADAPT-HMI maps between 01-Mar-2020 08:00 UT and 31-Jul-2020 08:00 UT with 12 hour cadence as input. This work used version 5.2 of the WSA model \citep{Kim19}}. Two of the WSA maps ({at 2020 April 17 08:00 UT (top panel) and at 2020 April 20 08:00 UT (bottom panel)}) of radial SW speed from the time series used to specify the inner BCs {are} shown in Figure~\ref{IBC} for latitudes between $-90^\circ$ and $90^\circ$. The Stonyhurst heliographic longitudes are in the range between $0^\circ$ and $360^\circ$. The projection of Earth lies on the central meridian and is marked in the plot. {It should be noted that Stonyhurst longitudes conventionally go from $-180^\circ$ to $+180^\circ$, with the central meridian at $0^\circ$. However, the ADAPT-HMI maps provided by the National Solar Observatory (NSO) in the ``Central Meridian Centered'' frame assumes the longitudes to go from $0^\circ$ to $360^\circ$ with the central meridian at $180^\circ$. The SW density and temperature at 0.1 au are estimated based on the assumptions of constant momentum flux and thermal pressure balance, with the free parameters $V_{fast}$, $V_{slow}$, $N_{fast}$, and $T_{fast}$ set to 700 km s$^{-1}$, 200 km s$^{-1}$, 200 cm$^{-3}$, and 2 MK, respectively \citep[e.g.,][]{Jian16}. The WSA magnetic field is scaled uniformly by a factor of 2 to compensate for the systemic underestimation of magnetic field strength at Earth \citep[e.g.,][]{Kim19}. The time series of WSA maps used in this study is not available publicly. However the ADAPT-HMI maps in both the Carrington and Central Meridian Centered frames are available through NSO website at \url{https://gong.nso.edu/adapt/maps/special/psp/adapt_hmi-los/}.} By solving the MHD equations with the BCs based on the time series of WSA maps, we obtain a data-driven, time-dependent SW background. {For {the} CME simulation, we used the ADAPT-WSA realization R006 {since it} best reproduced the SW background at Earth over a full solar rotation ($\sim$ 27 days) centered around the CME eruption time.}

Once the ambient  SW flow is determined, a flux rope is inserted over it, as described in \citet{Singh22}. Note that the SW magnetic field and velocity are replaced with those of the flux rope during this {insertion process}, whereas the flux rope density and {magnetic} energy density are added to the SW density and {total} energy density, respectively. {The magnetic field inside the flux rope is specified by the uniform twist solution given by \cite{Vandas17} as follows:
$$B_\mathrm{r} = 0; \\
B_\phi = \frac{B_\mathrm{0}}{1+b^2r^2};\\
B_\theta = -\frac{B_\mathrm{0}R_\mathrm{0}br}{(1+b^2r^2)(R_\mathrm{0}+r\cos\theta)}$$
Here $r$, $\theta$, and $\phi$ are the coordinates of a toroidal coordinate system \citep{Vandas17}. $B_\mathrm{0}$ and  $b$ are used to specify the magnetic flux and the number of magnetic field line turns, respectively. Though this solution is valid for a torus shape of major radius $R_\mathrm{0}$, \citet{Singh22} showed that it can also be used to prescribe magnetic field inside a flux rope shape defined by \citet{Isavnin16} by assuming it to be locally toroidal. The flux rope shape is defined by the following equations:
$$R(\phi) = \frac{R_\mathrm{p}}{R_\mathrm{t}}r(\phi);\quad r(\phi) = R_\mathrm{t}\textup{cos}^n\left(\frac{\pi}{2}\frac{\phi}{\phi_\mathrm{hw}}\right)$$
}
{Here $R(\phi)$ is the radius of the cross-section of the flux rope and $r(\phi)$ is the distance of its curved axis from the origin, $n$ is a free parameter which can be used to adjust the flatness of the flux rope shape, and $\phi_\mathrm{hw}$ describes the half angle of the flux rope. Therefore, to describe the shape and the magnetic field of the flux rope, following parameters need to be prescribed: $R_p$, $R_t$, $n$, $\phi_\mathrm{hw}$, $B_\mathrm{0}$, $R_\mathrm{0}$, and $b$.}

{At the time of CME insertion, we replace the magnetic field of the SW background with the magnetic field of the flux rope. Though the magnetic field divergence in the solar wind background is eliminated using the Powell method \citep{Powell99}, and the initialized analytic flux rope has zero magnetic divergence by design, there may appear a non-zero magnetic divergence {especially} at the interface of the inserted flux rope and the SW. This magnetic divergence is non-physical which may cause the simulation to become numerically unstable. However, we have observed that the divergence created during {the} flux rope insertion does not have any impact of the stability of the simulation. This was further confirmed by \citep{Singh23}, where they performed 456 CME simulations with this method without experiencing any magnetic divergence related instabilities. Therefore, we conclude that the Powell method is able to propagate the non-zero divergence successfully out of the computational domain.} {In the future, we will study the effect of magnetic divergence on the simulation results by performing similar simulations with other divergence cleaning approaches such as that suggested by \citet{Dedner02}.}

In this study, we assumed the total mass in the flux rope to be $10^{10}$ kg, which is two orders of magnitude less than the typical mass of a CME. This mass is uniformly distributed in the flux rope {with a number density of 27.3 cm$^{-3}$}. The rationale behind choosing a relatively minimal mass for the flux rope stems from observations that in CMEs, the flux rope usually comprises only a minor portion of the CME's overall mass. This is supported by the observation of a dark cavity within the standard three-part structure of a CME, indicative of the flux rope's position. {During the flux rope insertion, the total energy density in the flux rope volume is set as} 

$$
e_{total} = \frac{p_{SW}}{\gamma - 1} + \frac{|\mathbf{b}_{SW}|^2 + |\mathbf{b}_{FR}|^2}{8\pi} + \frac{\rho_{SW} |\mathbf{v}_{SW}|^2}{2},
$$
{where $\rho$, $p$, $\mathbf{v}$, and $\mathbf{b}$ are the density, thermal pressure, bulk velocity, and magnetic field, respectively. We have kept the adiabatic index $\gamma=1.5$ in this study. This {ad-hoc} definition of ${e_{total}}$ ensures a very small plasma $\beta$ inside the flux rope \citep{Singh20b}.}

As discussed in the Section~\ref{subsec:Constrain}, the CME properties are found using the GCS model. It is also worth noting that the CME height was $10R_\odot$ when it exited the LASCO C2 FOV, which is the last time the GCS fit was made. However, we insert a fully-formed flux rope into the IH model, i.e., when its apex height is $70R_\odot$. While doing this, we assume a self-similar CME expansion from $10R_\odot$ to $70R_\odot$. Using the drag-based model (DBM) \citep{Vrsnak07}, we estimated that the CME should reach the apex height of $70R_\odot$ on {2020 April 17} 08:39 with a speed of about 293 km s$^{-1}$. While using the DBM model, we adopt the drag parameter of $0.1\times10^{-7}$ and the asymptotic SW speed of 450 km s$^{-1}$. {These drag parameters are chosen to} ensure roughly the same arrival time accuracy for the DBM and WSA-ENLIL-Cone models \citep{Vrsnak14}. The flux rope is inserted using this DBM time and speed, but with direction, tilt, half angle, and aspect ratio given by the GCS model. {Following \citet{Singh22}, the initial velocity at any location inside the flux rope is defined as a combination of the radial and expansion velocities, i.e., $\mathbf{V} = \mathbf{V}_\mathrm{rad} + \mathbf{V}_\mathrm{exp}$, where $\mathbf{V}_\mathrm{rad}$ is in the radial direction away from the solar center and $\mathbf{V}_\mathrm{exp}$ is in the direction pointing away from the curved inner axis of the flux rope. If $V_\mathrm{CME}$ is the CME speed at apex, the following magnitudes of $\mathbf{V}_\mathrm{rad}$ and $\mathbf{V}_\mathrm{exp}$ ensure a self similarly expanding profile
\[
|V_\mathrm{rad}| = \frac{V_\mathrm{CME}}{1+R_\mathrm{p}/R_\mathrm{t}}, \quad |V_\mathrm{exp}(r_\mathrm{p})| = \frac{r_\mathrm{p}}{R_\mathrm{t}}|V_\mathrm{rad}|,
\]
where $r_\mathrm{p}$ is the radial coordinate inside the local torus structure.}

\begin{figure}
\centering
\center
\includegraphics[scale=0.1,angle=0,width=8cm,keepaspectratio]{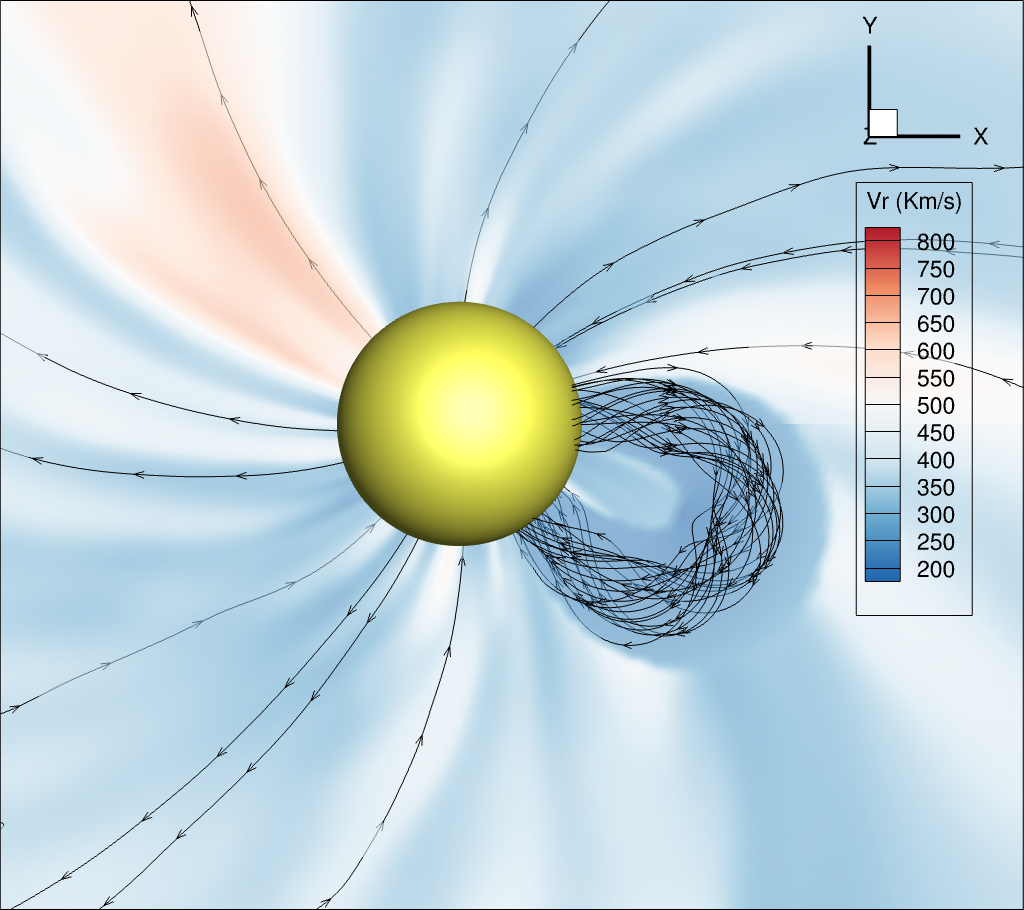}
\caption{The initial structure of a constant turn flux rope used to simulate the CME inserted into the IH at {2020 April 17} 08:39 with apex at $70R_\odot$ and SW  speed at it is 293 km s$^{-1}$. The poloidal and toroidal magnetic fluxes are set to $10\times10^{21}$ Mx and  $5.1\times10^{21}$ Mx, respectively. Its latitude, longitude, and tilt are $4.5^\circ$, $25.5^\circ$, and $10.1^\circ$, respectively. The semi-translucent slice represents the solar equatorial plane and is colored with radial plasma velocity. Magnetic field lines are shown with black lines. The yellow heliocentric sphere with radius of  0.1 au shows the inner boundary of the IH model.}
\label{Initial_FR}
\end{figure}

Since this was a stealth CME, we were not able to determine the magnetic fluxes and the helicity sign using EUV and magnetogram observations, as it had been done in \cite{Singh22}. Therefore, we assumed a poloidal flux of $10 \times 10^{21}$ Mx. The toroidal flux was found from this poloidal flux using the empirical formula relating them \citep{Qiu07}. This resulted in the toroidal flux of $5.1 \times 10^{21}$ Mx. The helicity sign was determined using the ``hemispheric helicity rule'' that states that flux ropes originating from the northern (southern) hemisphere of the Sun have negative (positive) helicity sign \citep{Pevtsov03}. \citet{Pevtsov14} showed that this simple assumption holds for 60-75\% of the flux ropes. As will be seen later from our simulation results, these  assumptions are very reasonable. The flux rope at the time of insertion is shown in Figure~\ref{Initial_FR}. {The flux rope parameters that prescribe the desired magnetic fluxes and ensure the shape resembling the fitted GCS shape are $R_p = 14.9 R_\odot$, $R_t = 55.1 R_\odot$, $n = 0.22$, $\phi_\mathrm{hw} = 25.8^\circ$, $B_\mathrm{0} = 0.0031$ G, $R_\mathrm{0} = 14.4 R_\odot$, and $b = 23.4$ au$^{-1}$.} The flux rope features the winding magnetic field lines with two legs on the inner boundary. The solar equatorial plane slice is colored according to the plasma radial velocity component. One can observe that the speed with which we initialize the CME is lower than that of the SW  surrounding it.

\begin{figure}
\centering
\includegraphics[scale=0.1,angle=0,width=6.7cm,keepaspectratio]{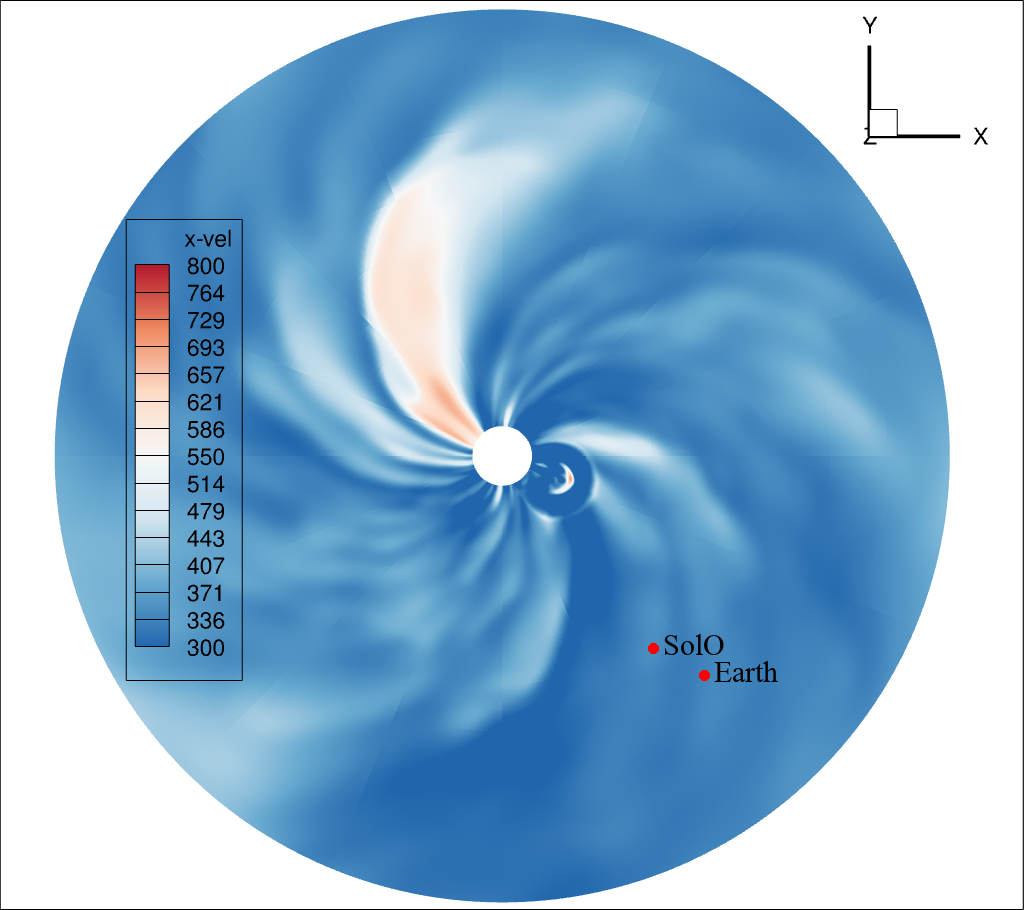}
\put(-144,140){\color{black}2020 April 17 09:58}
\includegraphics[scale=0.1,angle=0,width=6.7cm,keepaspectratio]{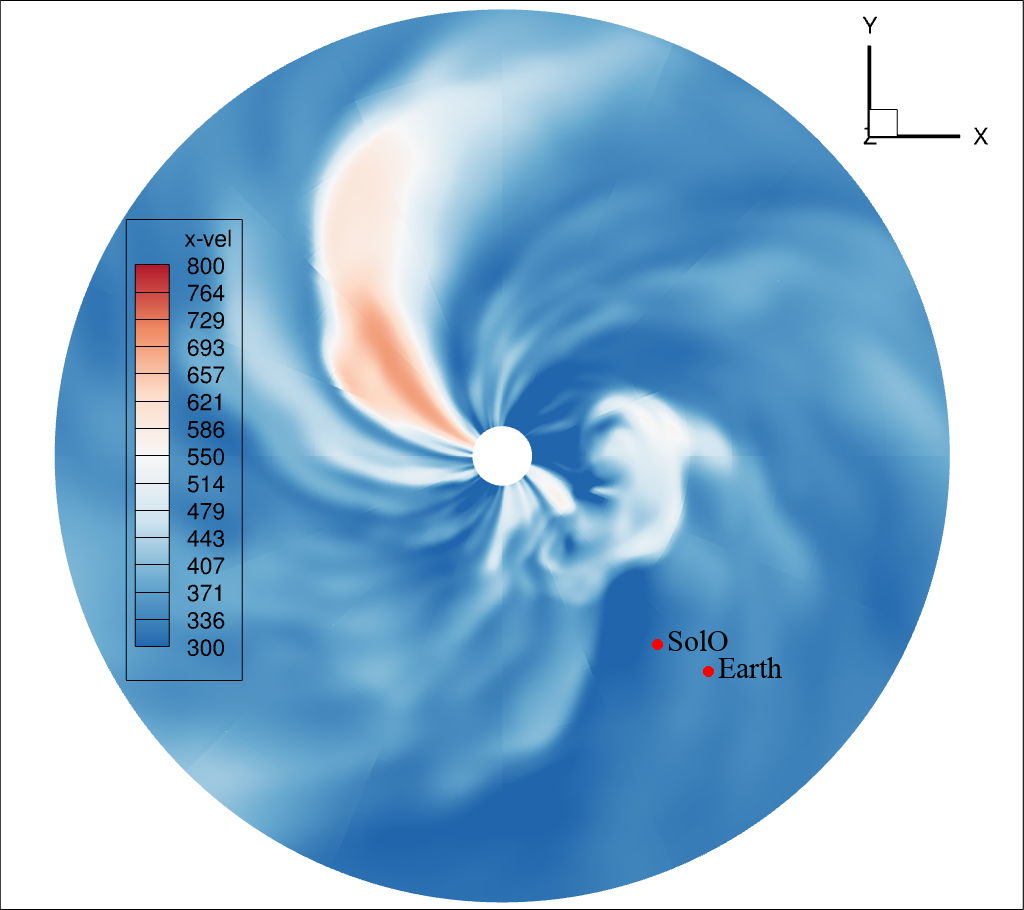} 
\put(-144,140){\color{black}2020 April 18 11:07}\\ 
\includegraphics[scale=0.1,angle=0,width=6.7cm,keepaspectratio]{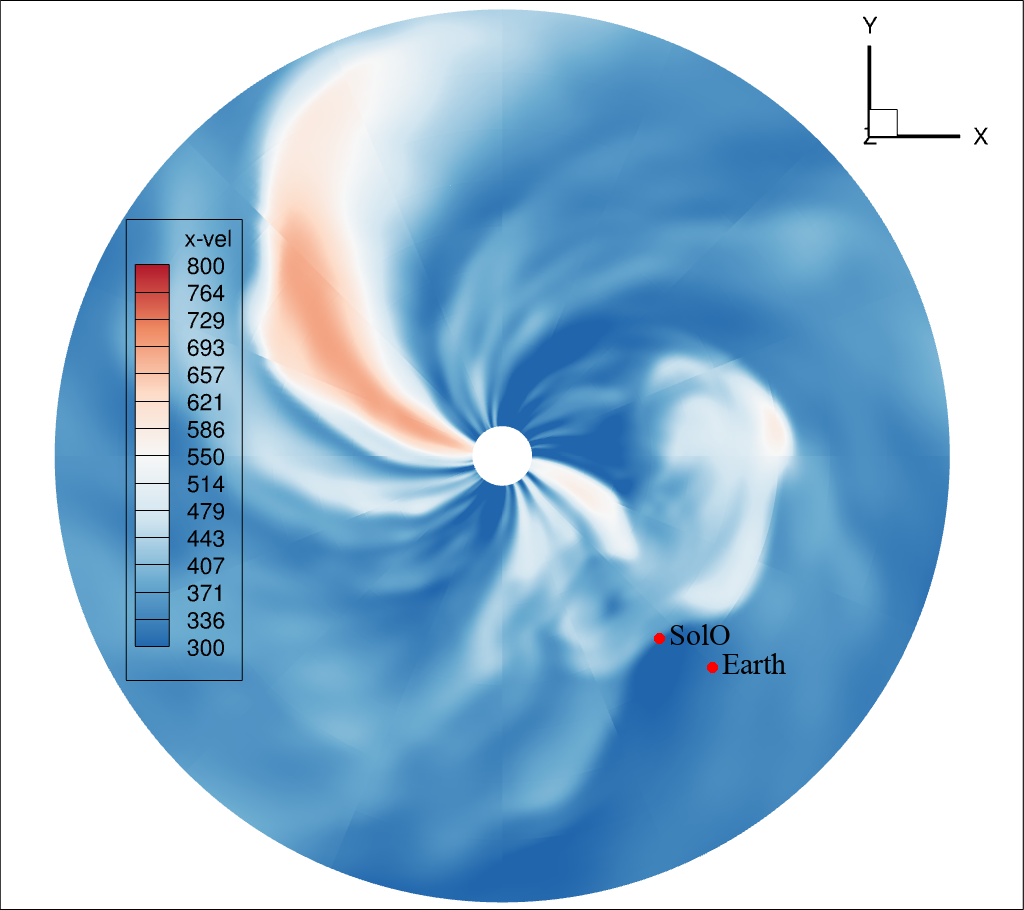}
\put(-144,140){\color{black}2020 April 19 12:03}
\includegraphics[scale=0.1,angle=0,width=6.7cm,keepaspectratio]{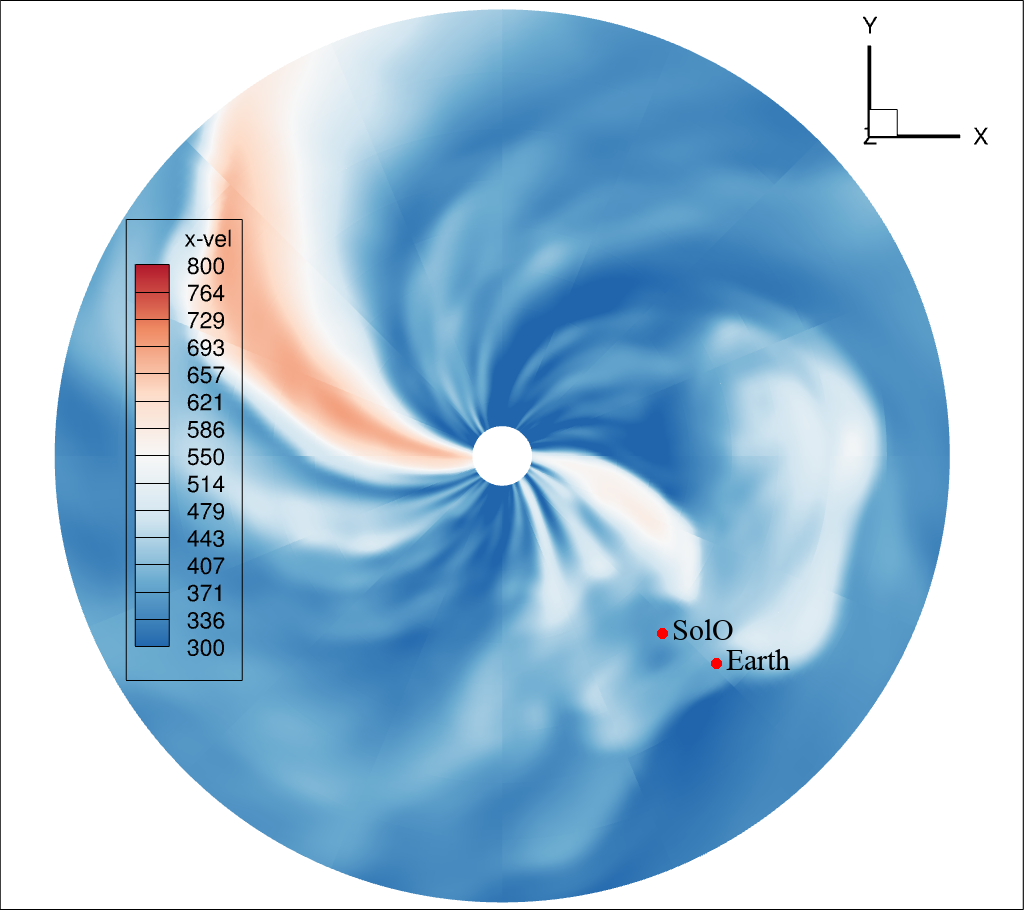}
\put(-144,140){\color{black}2020 April 20 12:28}
\put(-85,60){\color{black}\vector(1,1){10}}
\caption{The simulation results shown on {2020 April 17} 09:58 (\textit{top-left}), {2020 April 18} 11:07 (\textit{top-right}), {2020 April 19} 12:03 (\textit{bottom-left}), and {2020 April 20} 12:28 (\textit{bottom-right}). Each panel shows the radial velocity component in the solar equatorial plane. The location of SolO and Earth projected onto the plane are marked by red circles. The high-speed SW stream behind the CME is marked with the black arrow in the bottom-right panel. The inner and outer boundaries of the simulation domain are at $R=0.1$~au and $R=1.5$~au, respectively.}
\label{in-IH}
\end{figure}

Figure~\ref{in-IH} shows the propagation of the simulated CME through the IH starting from {close to} the insertion time until it passes the Earth. Each panel shows the radial velocity distributions in the solar equatorial plane. One can see that although the initial CME speed is slower than in the ambient SW (see also Figure~\ref{Initial_FR}), the former eventually exceeds the latter. This is likely because the CME entered the fast SW, which is especially well seen in the region behind it, marked with a black arrow in the bottom-right panel of Figure~\ref{in-IH}. 
{Another factor that can contribute to the CME acceleration is the pressure imbalance between  the inserted flux rope and the background solar wind, {which 
depends on
how the magnetic field, internal energy and plasma flow are chosen inside the CME and SW.} For the CME considered in this work, elevated densities observed in the rear CME side  at Earth and SolO (discussed below) indicate that the acceleration in this case is likely due to the fast SW pushing the CME.}
One can also notice that the CME front deviates from its initial smooth shape and becomes more complex due to the CME interaction with the background SW.

\begin{figure}
\centering
\center
\includegraphics[scale=0.1,angle=0,width=6.7cm,keepaspectratio]{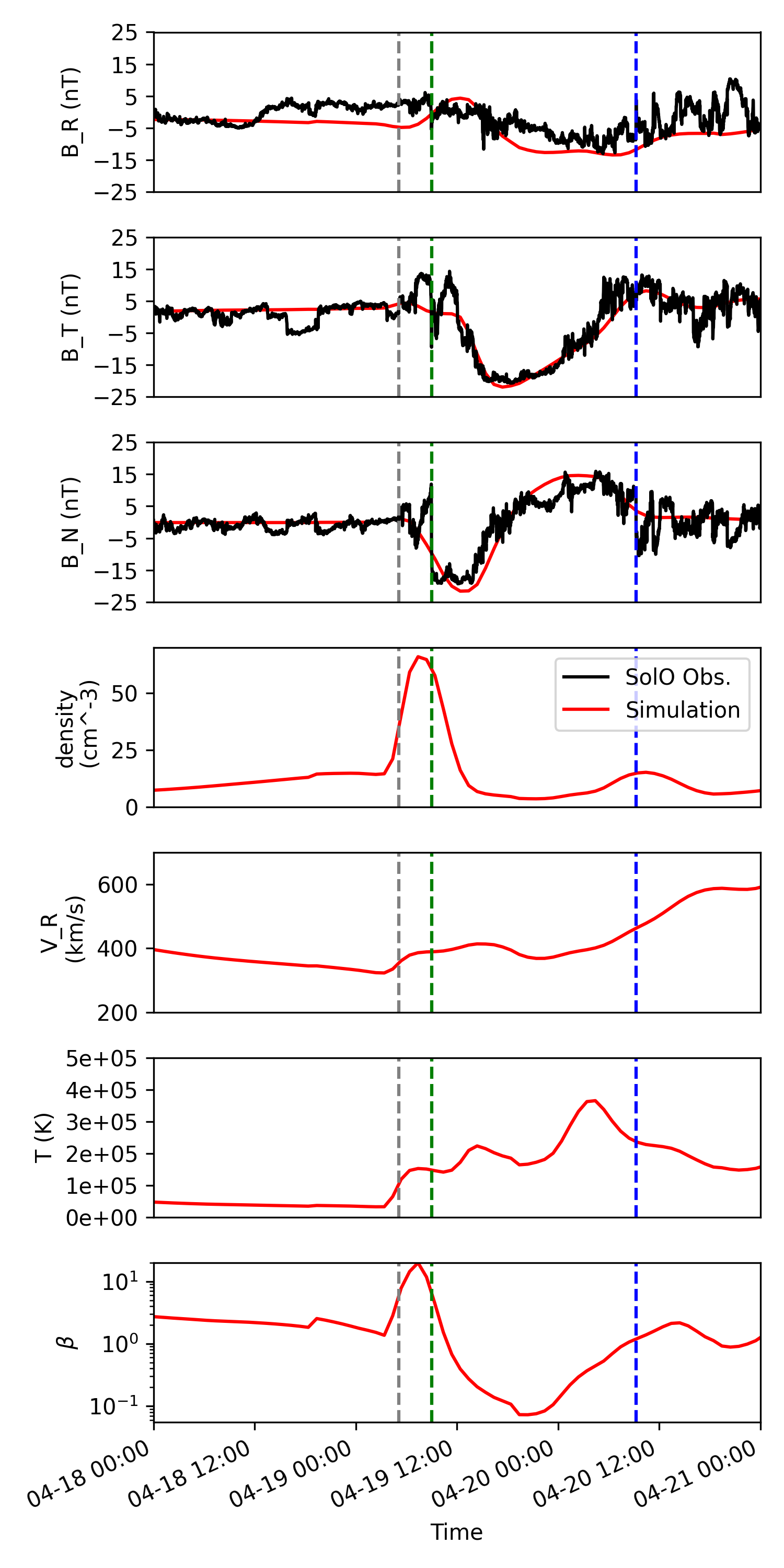}
\includegraphics[scale=0.1,angle=0,width=6.7cm,keepaspectratio]{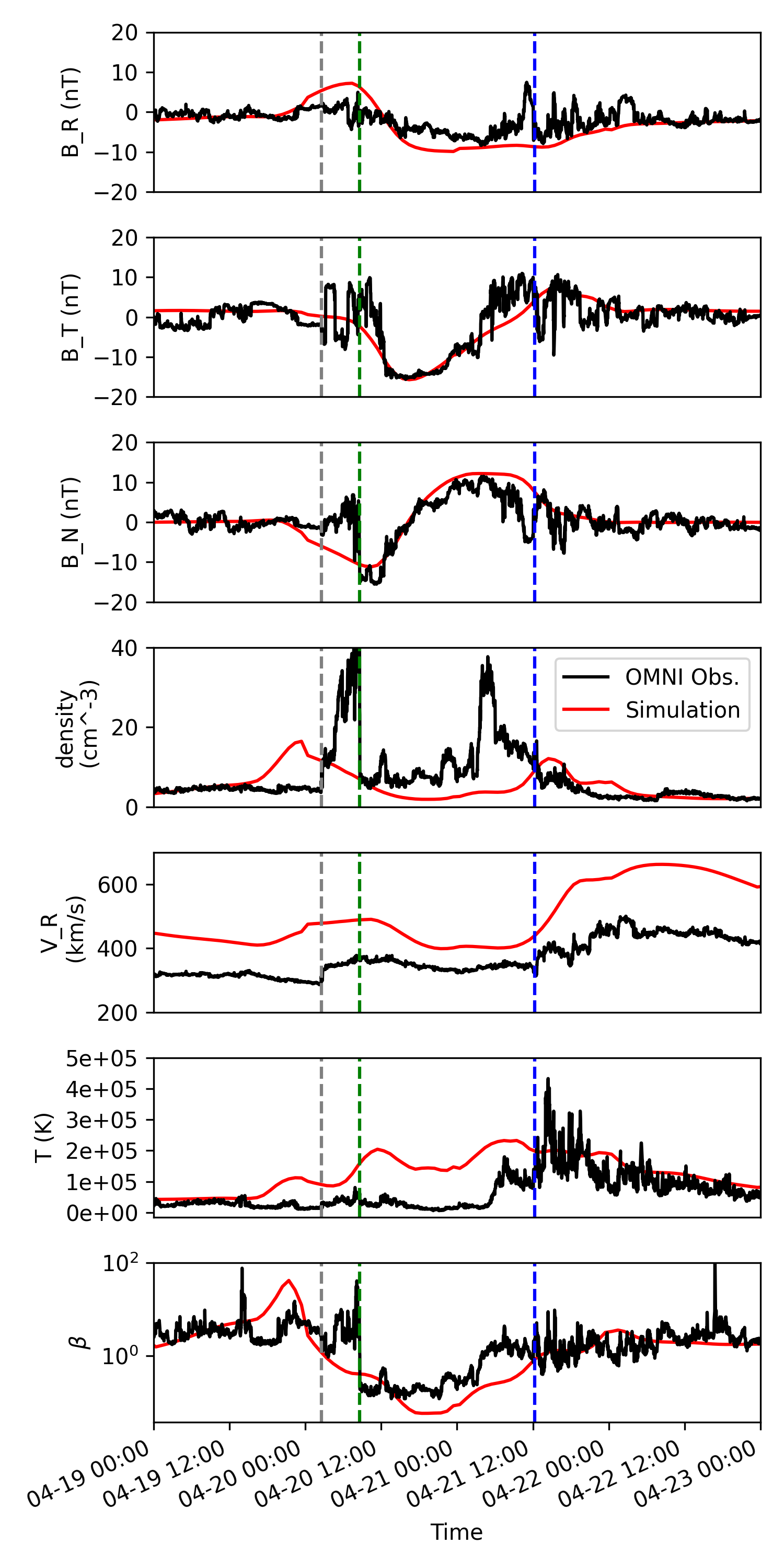}
\caption{The in situ, 1-min averaged, plasma and magnetic field properties of the chosen {I}CME, as well as the simulation results extracted hourly, are shown at SolO (\textit{left panel}) and at Earth (OMNI data) (\textit{right panel}). The observations and simulation results are shown with black and red lines, respectively. The SolO plasma data were unavailable at that time. To ensure a better magnetic field comparison in the simulation and observations, the time-series of simulation results at SolO had to be shifted forward by 5 hours. No such shift is done at Earth. The shock arrivals are marked with the {gray} dashed vertical lines. {The start and end of the smoothly rotating magnetic field are marked with the green and blue dashed vertical lines.}}
\label{in-situ}
\end{figure}

In figure~\ref{in-situ}, we compare our simulation results with {in situ} observations. This is done by extracting the simulation results along the trajectories of SolO and Earth every hour, and comparing them with the 1-min averaged SolO and OMNI data. This comparison is done for the magnetic field components in radial-tangential-normal (RTN) coordinates, as well as for plasma density, radial velocity{, temperature, and plasma $\beta$}. During the time of the CME passing over SolO, the Solar Wind Analyzer \citep[SWA,][]{Owen20} instrument onboard had not been fully commissioned. Therefore, the {plasma} data is unavailable at SolO, so this comparison was only possible at Earth. The left and right panels 
of Figure~\ref{in-situ} compare the simulation results with the data  at SolO and at Earth, respectively. The shock arrival, {start and end of smoothly rotating magnetic field} are marked, respectively, with {gray}, green, and blue dashed vertical lines.

The {I}CME arrived at SolO with a smoothly rotating {magnetic field} and a weak shock in front of it. In the absence of plasma data, the shock arrival is deduced from the time of a small jump in $B_N$. The shock arrival time at SolO, as derived from the 1-min cadence data, was {2020 April 19} 05:06. The {smooth rotation of the magnetic field vector starts on 2020 April 19 08:59 and ends on 2020 April 20 09:15.} The {start of the smooth rotation of the magnetic field} was identified by a sharp drop in $B_N$. The {identification of the rear boundary of that region is somewhat} subjective. The simulated {I}CME arrived at SolO later than the observed one. For this reason, we shift the simulation data (red lines) in the left panel ahead by 5 hours to allow for a better comparison between the simulated and observed magnetic field in the {ICME}. The simulated {I}CME does not show any sudden jumps in the magnetic field components at the shock arrival, but one can see a large density enhancement and a small $V_R$ increase. There is a good qualitative agreement in the magnetic field components in the {ICME}, especially in the $B_T$ and $B_N$ components. The simulation matches the observations not only in the sign of these magnetic field components, but also in their magnitude. The $B_R$ component is significantly smaller than the other components in the observed {I}CME, a behaviour well reproduced by our simulations.

In the right panel of Figure~\ref{in-situ}, the simulation results and observations are compared at Earth. The observed shock arrival time was {2020 April 20} 02:34. This time was identified by the sharp jumps in density and speed. The smooth rotation of magnetic field components {begins on 2020 April at 20 08:36 and ends on 2020 April 21 at 12:18}. For this comparison, no time shift was made contrary to what was done for SolO. Unlike {the in situ} observations, the simulated {I}CME did not reproduce the sharp jumps in density and speed, likely due to the insufficient grid resolution. Instead of reproducing the sudden enhancements in density and speed, a gradual rise  is seen, which starts $\sim 5$ hours earlier than the observed shock arrival. Regardless of that, the magnetic field profiles agree better without any time shifting. This is because the high density region in front of the simulated {ICME} appears to have a slightly larger duration than the observed sheath region. {We also notice that this density enhancement is about 4 times stronger  at SolO compared to Earth. For a radially expanding structure, one would expect the quantity of $\rho r^2 V \Delta t$ to be preserved along two radially aligned locations, where $\rho$ is the density, $r$ is the distance, $V$ is the speed, and $\Delta t$ is the duration for which the structure was present in the measurements. SolO, at 0.81 au, observed the density enhancement for 10 hours with average speed of 370 km s$^{-1}$. At 1 au, the same enhancement was observed for 19 hours with average speed of 480 km s$^{-1}$. For the above mentioned quantity to be same at SolO and Earth, we expect the density at SolO to be $\sim$ 4 times higher than at Earth, which is well produced in the simulation.}

Similarly to SolO, the signs and absolute values of $B_T$ and $B_N$ show a remarkable agreement at Earth. Likewise, the $B_R$ component is smaller than the other components both in the observations and simulations.  The regions of large density reveal themselves in the rear of the observed {ICME}. The simulated {ICME} shows a similar behavior, but the density enhancement is much smaller than the observed values. In the observed {ICME}, the density enhancement coincides with the rear portion of the {rotating magnetic field region}. In the simulation, the density enhancement occurs at the {end of magnetic field rotation}. {The simulation is able to reproduce a low plasma $\beta$ region inside the ICME. However, the simulated temperature profile inside the ICME is significantly higher than in the observations. This is in line with the similar temperature enhancements obtained in the other flux rope based simulations \citep[e.g.][]{Shen14, Mayank24}. Further research needs to be done to understand this unrealistically high heating of the plasma inside the ICME, so that more realistic thermodynamic evolution of the ICMEs can be obtained.}

In the right panel of figure~\ref{in-situ} the simulated SW appears to be much faster that in the observations, both in front of and behind the {I}CME. Consequently, the {I}CME speed is also higher in the simulation. This is the reason for its lagging behind the observations at SolO and being ahead of the observations at Earth. The reason for the simulated SW speed being faster is due to inner boundary conditions provided by the WSA model for the radial velocity component, as shown in Figure~\ref{IBC}. The higher speeds from the polar regions appear to extend to very low latitudes, which creates a very narrow and flat region of slow SW speed in {the} equatorial plane. Those high-speed regions extend as far as to Earth at low latitudes. Since the CME under consideration is very slow, the SW speed plays a major role in controlling its speed in the IH. The CME gets accelerated to much higher speeds than it was observed because of inaccuracies in the boundary conditions  provided by the WSA model. In fact, the discrepancies between simulated and observed CME speeds should have resulted in much larger errors in arrival time both at SolO and at Earth. In the next subsection, we explain the reasons of much smaller deviations.

\subsection{Comparison with HI data}
\begin{figure}
\centering
\center
\begin{tabular}{c c}
\includegraphics[scale=0.1,angle=90,width=9.4cm,keepaspectratio]{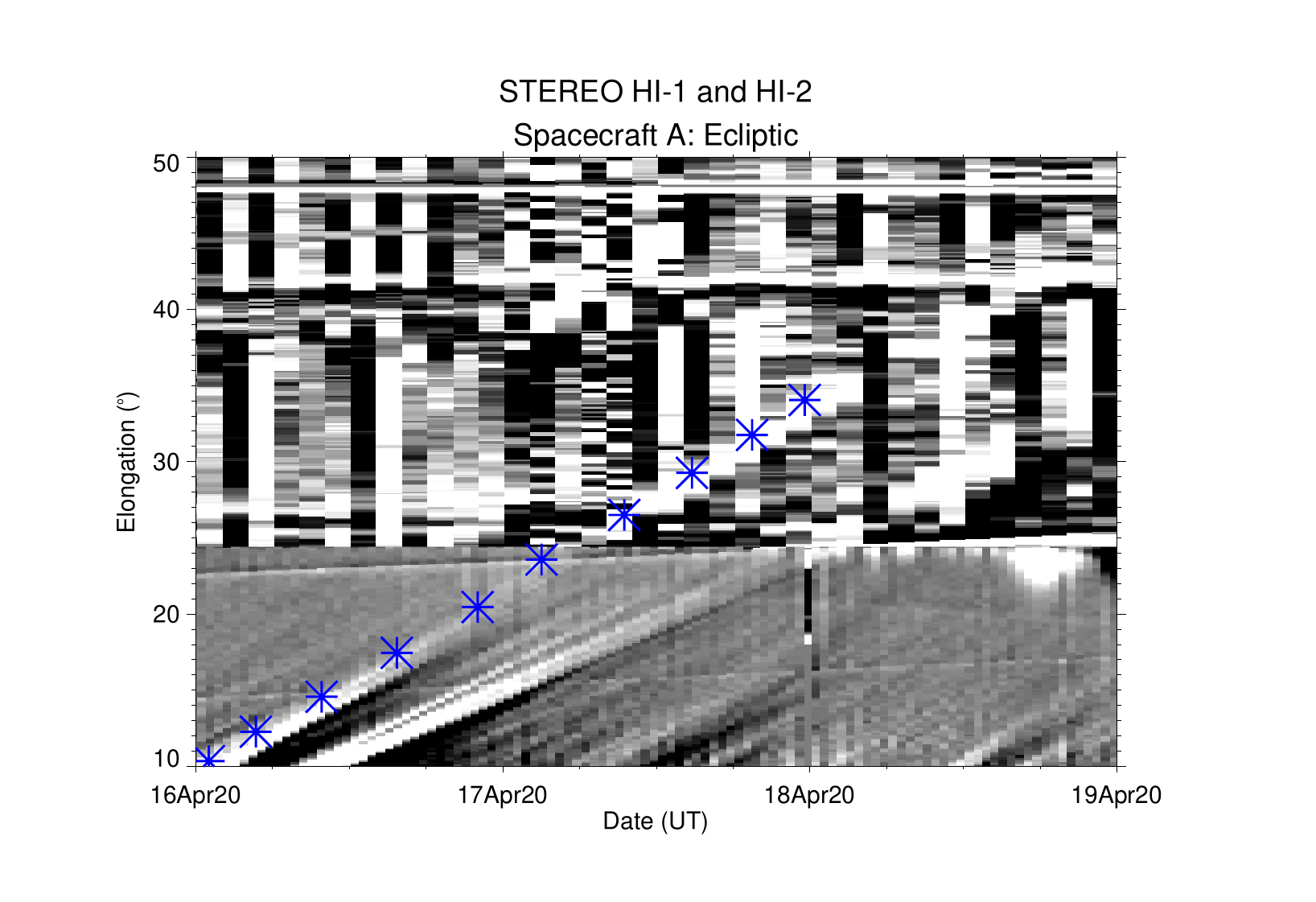}
\hspace{-1cm}
\includegraphics[scale=0.1,angle=0,width=8.5cm,height=6.1cm]{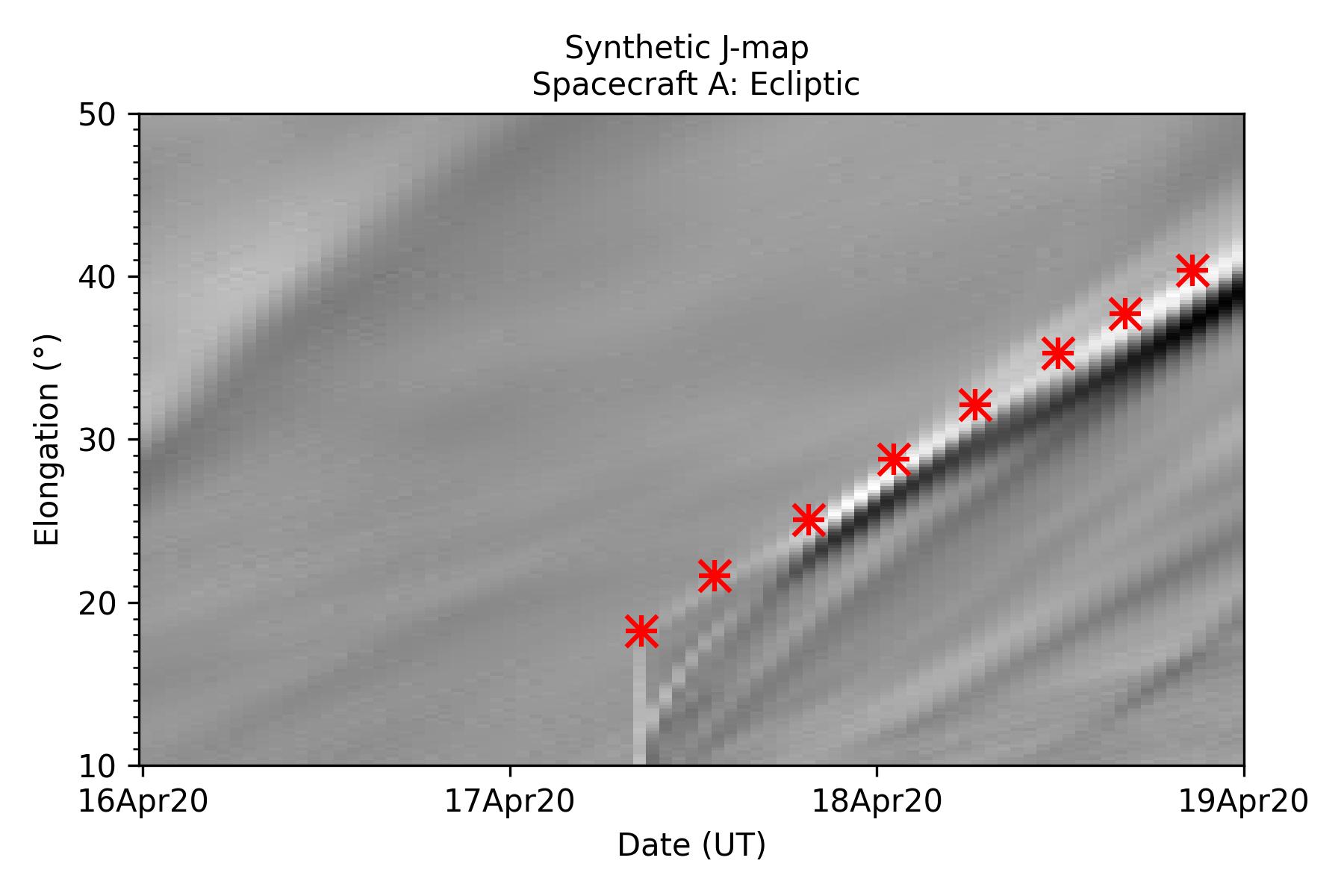}
\end{tabular}
\includegraphics[scale=0.1,angle=0,width=9cm,keepaspectratio]{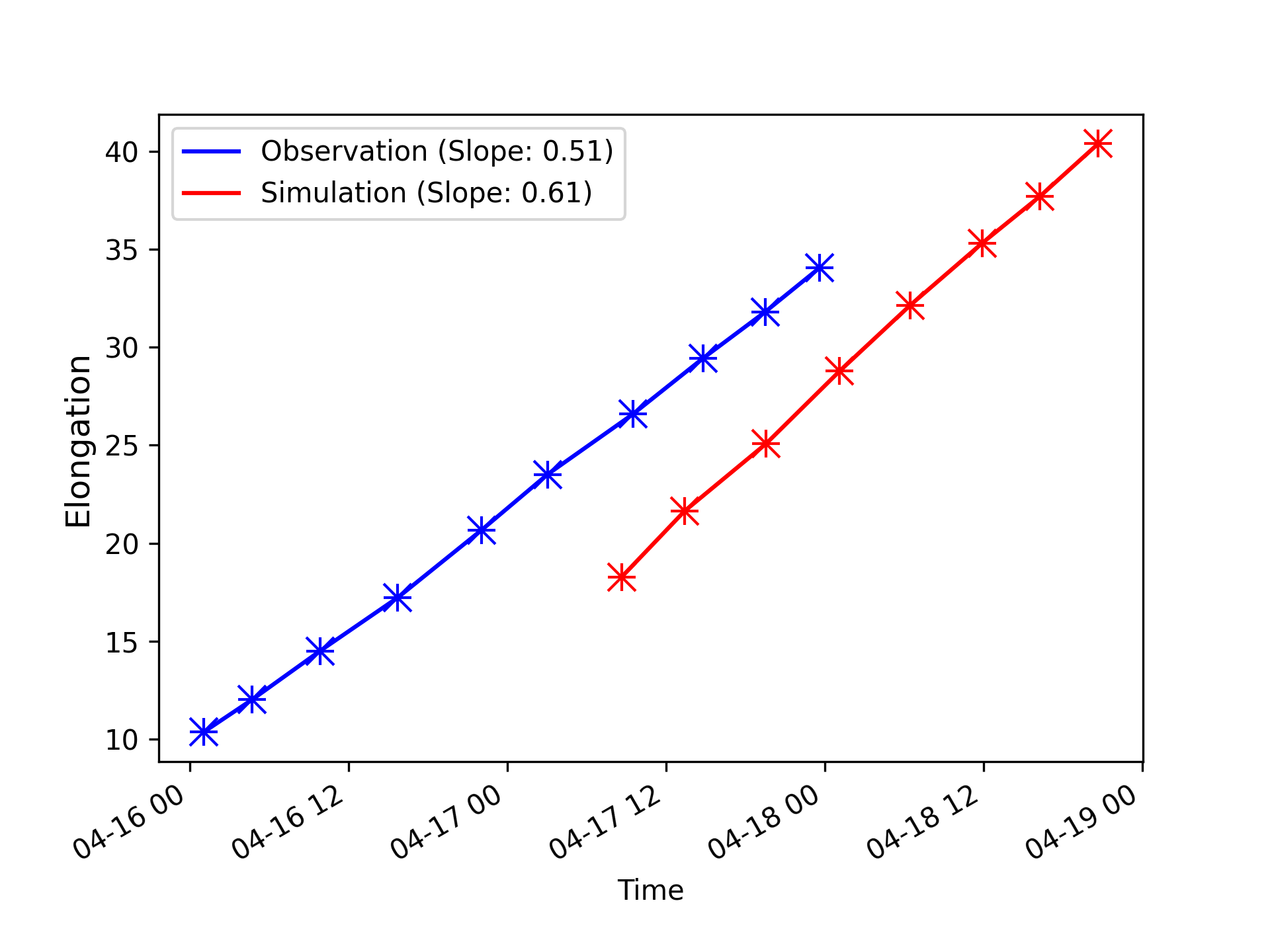}
\caption{(\textit{Top left panel:}) A J-map created with the HI-1 and HI-2 data from STEREO A, showing the CME as a diagonal bright line marked with blue stars. (\textit{Top right panel:}) A synthetic J-map showing the simulated CME in the STEREO A HI FOV. The bright front of the CME is marked with red stars. (\textit{Bottom panel:}) The CME fronts tracked in the observed (\textit{blue}) and synthetic (\textit{red}) J-maps. In the insert, the line slopes are shown, as estimated using a linear regression of elongation vs. the time- since-eruption data. Note that the simulated CME propagates faster than the observed one.}
\label{Jmaps}
\end{figure}

Here we compare our simulations with STEREO HI data in the IH. The HI1 imager on STEREO-A has a fixed Field of View (FOV) spanning from {$4^\circ$ to $24^\circ$} in terms of elongation angle. Likewise, the HI2 FOV ranges from $18.^\circ 7$ to $88.^\circ 7$ degrees in elongation angles. This is equivalent to covering distances up to and beyond 1 au, when combining both HI1 and HI2. As CMEs propagate through the IH, they expand and their density diminishes, leading to a decrease in their brightness within the HI FOV. Consequently, most CMEs become undetectable as they pass through the HI2 FOV. A prevalent method for tracking CMEs in the HI FOVs involves a creation of time-elongation maps, commonly known as J-maps. These maps are generated by selecting the pixels in running-difference HI images aligned with the ecliptic plane and stacking them vertically over each chosen time interval. This is how the J-maps are obtained. CME fronts in these maps reveal themselves as slanted bright streaks. The top-left panel of Figure~\ref{Jmaps} shows a J-map created using STEREO A HI1 and HI2 data for three days starting on {2020 April 16}. The considered CME has a bright front indicated in the figure with blue-cross markers.

In the top-left panel of Figure~\ref{Jmaps}, we show a synthetic J-map obtained by using the density calculated in our simulation domain. To create such a J-map, we perform another simulation without a CME inserted into the simulation region, thus obtaining the corresponding SW background. Afterwards, we subtract the ambient density from the solution which contains the inserted CME. The resulting difference has non-zero values only in the regions of the domain {containing density modifications due to the modeled CME}. The density differences along the ecliptic-plane lines of sight, with elongation angles ranging between 10$^\circ$ and 50$^\circ$, are extracted, integrated, and stacked vertically to create a synthetic J-map. In the synthetic J-map shown in the top-right panel of Figure~\ref{Jmaps}, the CME is indicated with red crosses.

In the bottom panel of Figure~\ref{Jmaps}, we compare the time-elongation profiles of the observed and simulated CMEs. This plot is made of the blue and red stars shown in the top panels. One can clearly see that the simulated CME trails the observed CME starting immediately from the moment of insertion. The CME insertion time and its speed, when its apex was at 70$R_\odot$, were calculated using the DBM, as discussed in section~\ref{sec:Sim_Results}. At the time of insertion, the front of the simulated CME had an elongation angle of $18^\circ$. The J-map comparison reveals that the observed CME had achieved this elongation angle 12 hours before the simulated CME. Furthermore, we found the linear-regression slopes in the data and simulation to be equal to 0.51 degrees/hour and 0.61 degrees/hour, respectively. This indicates that the simulated CME is moving faster than the observed one, likely due to the SW being faster in the simulation than in reality. However, these two mutually opposite errors practically cancel off each other, resulting in a good arrival time prediction at Earth. This demonstrates how errors in different components of the simulation workflow can interact, sometimes leading to a favorable prediction of the arrival time.

\section{Conclusion}\label{sec:Conclusions}
In this study we have reported the results of our MHD simulation of a stealth CME which was observed at SolO and Earth when they were in near-radial alignment. This CME did not have an apparent lift-off signature in the low corona{l} EUV observation{s} and had a slow speed of 120 km s$^{-1}$ in the STEREO A FOV. The corresponding {I}CME was observed both at SolO and Earth and involved a weak shock and smoothly rotating {magnetic field} signature at both locations. \citet{Davies21} compared the magnetic field profile in this {ICME} at SolO and Earth, and concluded that all the components were well correlated. These multi-spacecraft observations in radial alignment present a unique opportunity to test the ability of a CME model to reproduce the spacecraft data. Besides reporting our simulation results, we also used both {in situ} and heliospheric imager data for a detailed comparison of the simulated CME with the observations. The conclusions of our study are as follows.
\begin{enumerate}
    \item The CTFR model is suitable for simulating CME flux ropes and can be fine-tuned with a variety of observed CME properties. These include the CME velocity, direction, tilt, aspect ratio, and half angle, as deduced by the GCS model. Furthermore, the CTFR model can be  initialized with the{ information of the} magnetic poloidal and toroidal {magnetic} fluxes, and the helicity sign of the magnetic field twist within the flux rope.
    \item By comparing the simulation results with in situ observations at SolO and Earth, it was found that our simulations accurately reproduced the structure of {{the} rotating magnetic field} associated with the chosen CME. The simulation effectively mirrored substantial rotations in the $B_T$ and $B_N$ components observed inside the {ICME}.
    \item As compared with data, the simulated {I}CME arrival was underestimated  by 5 hours at SolO and overestimated by 5 hours at Earth. The CME speed, as well as the ambient SW speed, turned out to be higher in the simulation {compared to what} was observed at Earth. We also observed {smaller enhancement of the density} in front and on the back of the simulated {ICME} as compared t{o t}hose in the observed {event}.
    \item The elevated speed of the simulated SW at Earth {is a result of the} higher SW velocities at lower latitudes provided by the WSA model.
    \item A comparison of the time evolution of simulated elongation angles pertinent to the chosen CME with that observed by STEREO A indicates that the CME was introduced into our model domain 12 hours later than it should have been. This comparison also revealed that the simulated CME was traveling through the IH faster than the observed one, likely due to faster speeds in the simulated ambient SW. This discrepancy explains its delayed arrival at SolO and premature arrival at Earth.
    \item The time to insert this CME into the ambient SW was estimated using the DBM model, which resulted in a 12-hour {deviation}. Nonetheless, had the DBM model accurately predicted the insertion time, our MHD simulation would have resulted in a significantly greater discrepancy in the arrival times at SolO and Earth. This means that inaccuracies in different components of a forecasting approach can sometimes inadvertently negate each other, leading to unexpectedly accurate outcomes.
\end{enumerate}

Our simulations correctly  reproduced the distribution of {rotating} magnetic field inside the considered {ICME} both at SolO and at Earth, which shows that the CTFR model can correctly capture the radial evolution {of the} magnetic field associated with this {I}CME. However, it should be understood that 
magnetic field patterns observed for this {I}CME at SolO and Earth were closely correlated  \citep{Davies21}. There are other {events} that exhibit significantly lower correlation, even for nearly radially aligned spacecraft  \citep[see, e.g.,][]{Regnault23}. We intend to conduct further research on such {I}CMEs to assess whether MHD simulations using flux ropes, such as CTFR, can accurately reproduce this lack of correlation.

\section*{Acknowledgments}
The authors of this work are grateful for the support provided by jointly by the National Science Foundation (NSF) and NASA SWQU grant 2028154. DVH is thankful for the support provided
by the NASA FINESST grant 80NSSC22K005. TKK acknowledges additional support from the NASA grant 80NNSC20K1453. NP was also supported, in part, by NSF-BSF grant 2010450 and NASA grants 80NSSC19K0075 and 80NSSC21K0004.  Supercomputing resources supporting this work were made available by the NASA High-End Computing (HEC) Program awards SMD-20-92772410 and SMD-21-44038581 through the NASA Advanced Supercomputing (NAS) Division at Ames Research Center, as well as by the NSF \href{https://access-ci.org/}{ACCESS} project MCA07S033 on the Purdue Anvil and SDSC Expanse systems. We also acknowledge the use of data from the NASA/GSFC Space Physics Data Facility's OMNIWeb for the SW and interplanetary magnetic field data used in this study. Additionally, we made use of SOHO and STEREO coronagraph data from https://lasco-www.nrl.navy.mil and stereo-ssc.nascom.nasa.gov respectively. Lastly, we used the data from the SDO EUV and magnetogram from http://jsoc.stanford.edu/ajax/exportdata.html.
This work has utilized data that has been produced through collaboration between the Air Force Research Laboratory (AFRL) and the National Solar Observatory.

\bibliographystyle{aasjournal}

\end{document}